\documentclass[letterpaper]{article}
\usepackage{isea}
\usepackage[pdftex]{graphicx}
\usepackage{times}
\usepackage{helvet}
\usepackage{courier}
\usepackage[numbers]{natbib}
\usepackage{xcolor}
\usepackage{float}
\usepackage{caption}
\title{``Benefit Game: Alien Seaweed Swarms''–\\Real-time Gamification of Digital Seaweed Ecology}

\author{
1\textsuperscript{st} Dan-Lu Fei\textsuperscript{1,2}, 2\textsuperscript{nd} Zi-Wei Wu\textsuperscript{1,2}, 3\textsuperscript{rd} Kang Zhang\textsuperscript{2,*}\\
\textsuperscript{1}Division of Emerging Interdisciplinary Areas, Academy of Interdisciplinary Studies, \\Hong Kong University of Science and Technology, Hong Kong SAR, China\\
\textsuperscript{2}Computational Media and Arts, Information Hub,\\Hong Kong University of Science and Technology (Guangzhou), Guangzhou, China\\{\textsuperscript{*}kzhangcma@hkust-gz.edu.cn}
\\
\newline
\newline
}
\begin{document} 
\maketitle

\begin{abstract}
``Benefit Game: Alien Seaweed Swarms'' combines artificial life art and interactive game with installation to explore the impact of human activity on fragile seaweed ecosystems. The project aims to promote ecological consciousness by creating a balance in digital seaweed ecologies. Inspired by the real species ``Laminaria saccharina,'' the author employs Procedural Content Generation via Machine Learning technology to generate variations of virtual seaweeds and symbiotic fungi. The audience can explore the consequences of human activities through gameplay and observe the ecosystem's feedback on the benefits and risks of seaweed aquaculture. This Benefit Game offers dynamic and real-time responsive artificial seaweed ecosystems for an interactive experience that enhances ecological consciousness.

\end{abstract}

Digital Ecologies, Interactions, Sustainability, Ecological Consciousness,  Virtual Environments, Environmental Game


\section{Introduction}



In recent years, there has been growing concern about the impact of human activity on the natural world, including marine aquaculture. Anthropogenic interference can affect the genetic structure of seaweed populations, and a cybernetic system exists among seaweed, marine fungi, valuable bioproducts, and human activity.	

Inspired by this cybernetic system, our work ``Benefit Game: Alien Seaweed Swarms'' aims to create digital ecologies using Procedural Content Generation via Machine Learning~(PCGML) to generate a virtual marine environment of seaweed swarms and symbiotic fungi. The audience interacts with valuable game tokens to affect the growth of seaweed swarms. A healthy seaweed ecosystem can produce more valuable game tokens as profits, but excessive harvest can lead to the extinction of the seaweed ecologies. At this point, the environment would indicate that the seaweed ecosystem needs more tokens and time to recover. The audience as players aim to find a sustainable balance between human activities and the potential benefits and risks of seaweed aquaculture.

\begin{figure*}[ht]
\includegraphics[width=\textwidth]{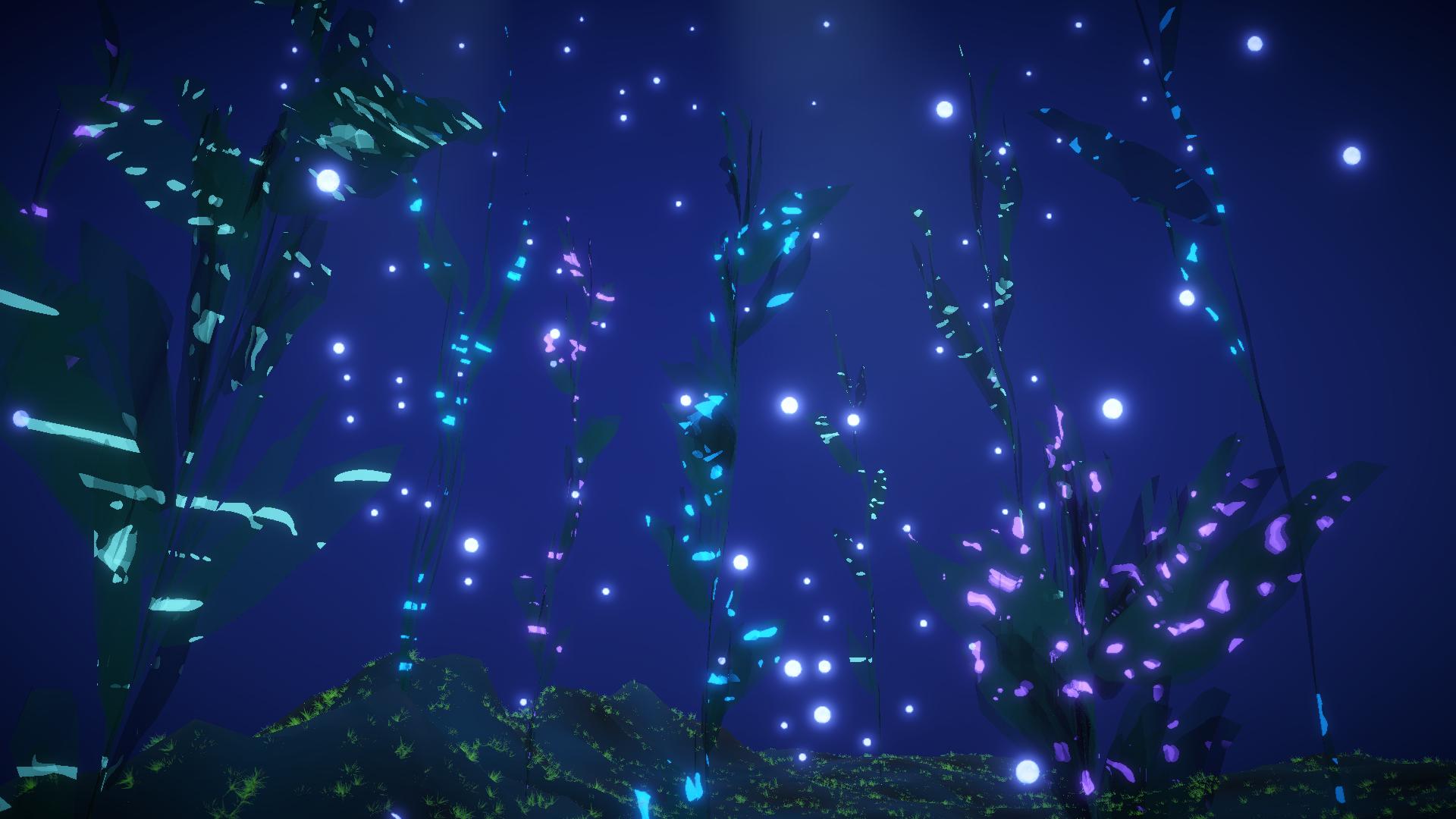}
\caption{A snapshot of digital seaweed ecology. \copyright Danlu Fei and Ziwei Wu.}
\label{fig:overall}
\end{figure*}

The contribution of our work includes the development of a virtual seaweed ecosystem generated in real-time using PCGML technology~(Fig.~\ref{fig:overall}), and an interactive gameplay installation~(Fig.~\ref{fig:installation}). Our work is an artistic game rather than simulating the reality. Ultimately, we hope to foster greater ecological consciousness and sustainability.	

\section{Background}
As Jason W. Moore argues in his book \textit{Anthropocene or Capitalocene: Nature, History, and the Crisis of Capitalism}~\cite{moore2016anthropocene} the rise of capitalism and its exploitation of natural resources has led to a fundamental transformation of the earth's ecosystems, creating what some refer to as the Anthropocene era. This era is characterized by the domination of humans over nature, with the pursuit of profit and economic growth taking precedence over environmental sustainability. \textit{A History of the World in Seven Cheap Things}~\cite{patel2017history} further argues that the exploitation of cheap nature, labor, and energy has been a driving force behind the expansion of capitalism and the colonization of new territories. ``Nature'' is not a thing but a way of organizing, representing the cheapening of life, yanked into processes of exchange and profit, denominated and controlled by capitalism. There is nothing like an ecological crisis to remind civilization that nature is never cheap. One should rethink the relationship between humans and nature.


Charles Robert Darwin~\cite{darwin1868variation} explored how artificial selection can influence the evolution of plants and animals. Recent research on seaweed~\cite{luttikhuizen2018strong} has further explored this idea, highlighting the impact of human colonization on the genetic structure of seaweed populations. Human activity in new territories can have significant consequences for the natural world, including changes to the genetic makeup of plant populations. These species are considered alien species in non-local areas under human activity. Laminaria saccharina~(known by the common name sugar kelp) is a representative species with commercial value, as it can be used as food, raw material for skincare products, and medicine. Research shows that this species' colonization history originated in the northwest Pacific, then crossed to the northeast Pacific in the Miocene. It also lived in the Arctic and northeast Atlantic, and postglacial recolonization led to its secondary contact in the Canadian Arctic~\cite{luttikhuizen2018strong}. This colonization history probably contributed to its use as ``Komb'' in Japanese cuisine, as well as its use as a raw material for American skincare companies. Owing to the economic value of this seaweed and its critical role as a habitat for various species, this seaweed has been introduced worldwide and is recognized as an alien species rather than an invasive one.

Cybernetics is concerned with ``control and communication in the animal and the machine~\cite{wiener2019cybernetics},'' and has been applied to a wide range of fields, including biology, engineering, and art. In the context of seaweed aquaculture, cybernetic approaches have been used to model and optimize the growth of seaweed populations, as well as to explore the potential for artificial propagation and colonization. 
In the cybernetic system of seaweed agriculture, there is a human link in the ``chain of transmission and return of information~\cite{wiener2019cybernetics},'' which is the chain of feedback~\cite{huang2021agricultural}. Humans observe seaweed production and develop aquaculture strategies in a timely manner. These cybernetic approaches have also been applied to the creation of computer-generated art. 

Against this backdrop, our artwork ``Benefit Game: Alien Seaweed Swarms'' seeks to engage the audience as the chain of feedback in computational cybernetic seaweed aquaculture, exploring the potential for a more balanced and sustainable coexistence with the environment. 

\section{Related Work}

\subsection{Environmental Sustainability}
The concept of environmental sustainability is crucial in the art field, and many artists are interested in designing sustainable systems. For example, \textit{PlantConnect}~\cite{PlantConnect} is a real-time artwork that explores human-plant interaction through a system that incorporates bioelectricity, light, sound, CO2, photosynthesis, and computational intelligence. This circuit enhances informational linkages between humans, plants, bacteria, and the physical environment.
\textit{Wayterways}~\cite{Waterways} creates a multi-channel audiovisual system to increase awareness of the fragile relationship between people and water in the Okanagan Valley.

Many video games also focus on ecology and sustainability. In \textit{ecoOcean}~\cite{EcoOcean}, four players are required to fish within the same ecosystem. If everyone catches as many fish as possible, the ecosystem will quickly deplete. Players need to communicate and adhere to common rules in order to achieve sustainable maximization of benefits. Similarly, games such as \textit{Working with Water} and \textit{Eco}~\cite{Working2021,Eco2018} aim to engage players in contemplating the balance between harvesting natural resources and protecting the ecosystem. Our game also encourages the audience to explore how to interact with the digital seaweed ecology in a sustainable way.

\subsection{Interactive Alife Artwork}
There have already been many existing artists and artworks in artificial life~\cite{wu2024survey} and games that discussed about ecology system. The most famous and classic one is \textit{Conway’s Game of Life}~\cite{Conway1970}. In the 1970s, John Horton Conway uses cellular automaton~(CA) to create Computational Ecosystems~(CEs)~\cite{Antunes2015}.
More recently,  influenced by the previous artificial life games such as \textit{SimCity} and \textit{The Sims}. Ian Cheng creates a series of works,  \textit{Emissaries}~\cite{Cheng2015} which he describes as a ``virtual ecosystem'' using modern AI technology. Besides the complex ecosystem, \textit{Mimicry}~\cite{Wu2021} by Ziwei Wu and Lingdong Huang explores the real-time texture generation of virtual insects based on a living environment. More conceptually, \textit{CMD: Experiments in Bio Algorithmic Politics}~\cite{Sedbon2019} by Michael Sedbon challenges the cybernetic forms of control between two artificial ecosystems sharing the same light source. \textit{Totem} is a dynamic art project that produces complex patterns in light using the self-developed software \textit{autonomX}~\cite{autonomX}. By utilizing the generator of \textit{autonomX} which is a grid-based wrapper for an AI and/or ALife algorithm, \textit{Totem} presents a performance for two human performers and eight spiking neural networks interacting with each other.
Our work builds a playable, real-time artificial life digital ecology and invites audience interaction through this dynamic system.

\subsection{PCG Game}
``Procedural Content Generation~(PCG) is the algorithmic creation of game content with limited or indirect user input''~\cite{Togelius2011}. PCG is extensively utilized in the generation of game assets. Eman et al. develops a game \textit{A Walk Alone}~\cite{Al-Zubeidi2022} that employs PCG for creating buildings, while Aaron Oldenburg integrates PCG in his art games~\cite{Oldenburg2018} to generate psychedelic landscapes that resemble everyday hallucinations, spiritual encounters, and so on.

Over the past decade, PCGML has gained momentum, primarily focusing on the generation of 2D game levels or sprites. Nonetheless, research has also been conducted on the creation of 3D models~\cite{Li2020,shi2020fast}. Our work employs PCGML to generate parameter-driven 3D seaweed models.



\section{Benefit Game: Alien Seaweed Swarms}
\subsection{Game System Design}
The essence of our gameplay is the interaction between the audience and the digital ecology we create, consisting of a Seaweed Swarm System and a symbiotic Fungi System~(Fig.~\ref{fig:system}).

\renewcommand{\thefigure}{3}
\begin{figure*}[t]
\centering
\includegraphics[width=5in]{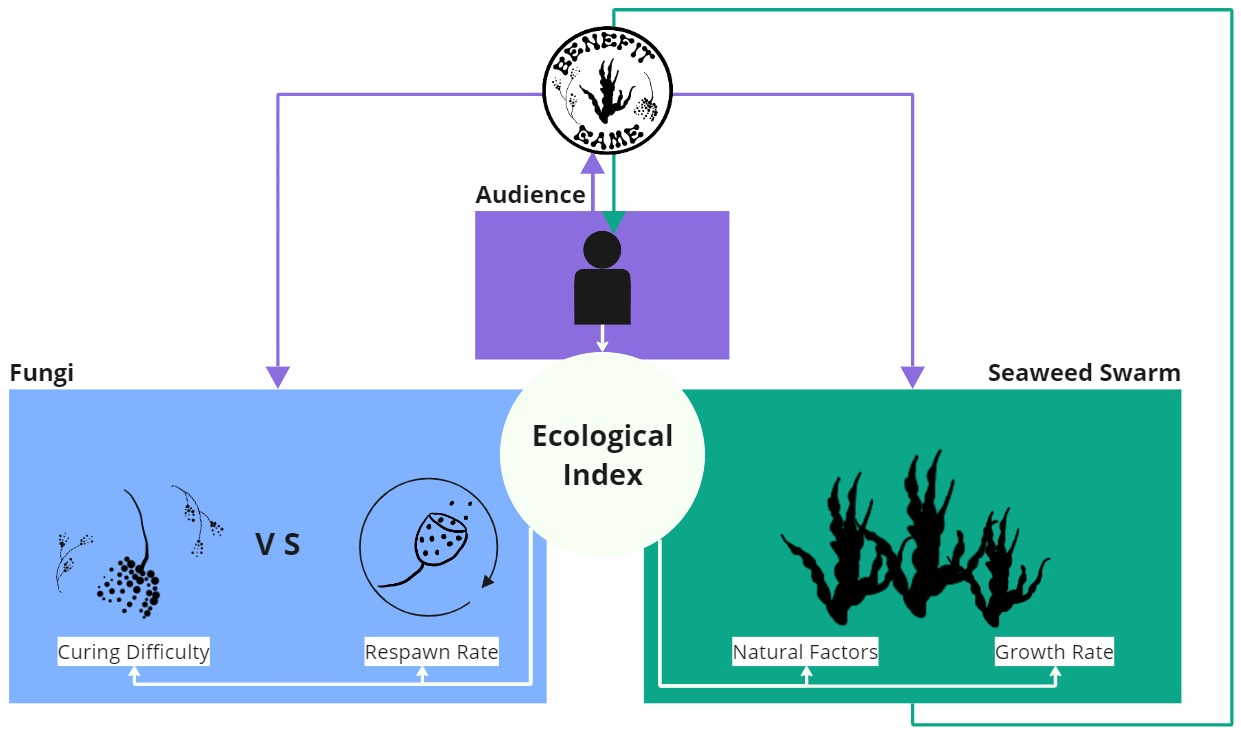}
\caption{Overall design of game system. \copyright Danlu Fei and Ziwei Wu.}
\label{fig:system}
\end{figure*}

\subsubsection{Seaweed Swarm System}
In real life, humans invest in seaweed aquaculture for profit. Similarly, in our game, the audience can harvest virtual seaweed by inserting game tokens. The harvested seaweed is periodically~(every 20 seconds) settled based on its selling price and returned to the audience as game tokens for further interaction. The price of virtual seaweed is determined by its shape, which is calculated based on the blade width, blade length, blade density, and stipe length of the harvested seaweed.
\subsubsection{Fungi System}
In the process of seaweed aquaculture, parasitic oomycetes can cause rot disease on the seaweed blades, leading to irregular rotted spots and reduced selling price~\cite{Vallet2018}. However, certain symbiotic fungi can kill the oomycetes and treat the disease~\cite{Tourneroche_2020,Patyshakuliyeva_2019,Pang_2010,Vallet2018}. 
Our game design includes a Fungi System where the audience can cultivate fungi by inserting game tokens, gradually improving the seaweed's health condition. The rotted spots on the seaweed blades will become smaller and less dense~(detailed in the Technical Implementation's Disease Effect section), and the seaweed's selling price will gradually increase. When the number of fungi is sufficient to kill the oomycetes, the seaweed swarm will recover completely. Oomycetes, however, will periodically respawn, causing the seaweed to become diseased again. The audience needs to balance between harvesting seaweed for profit and treating seaweed disease.

\subsubsection{Ecological Index}
We define a global variable, the ecological index~(EI) to represent the impact of the audience's interaction with the Seaweed Swarm System on the ecosystem. During seaweed aquaculture, human activities can impact the ecosystem positively or negatively, which in turn affects the Seaweed Swarm System and the Fungi System, including the growth rate of seaweed, the curing difficulty of rot disease, and the respawn rate of pathogenic oomycetes. A higher EI indicates a better ecological condition, while a negative EI represents ecological crisis. As shown in Fig.~\ref{fig:ei} The relationship between EI and the audience's token insertions for the Seaweed Swarm System is a periodic function consisting of three sine functions, corresponding to the ``prosperity of growth'' stage, the ``decline of growth'' stage, and the ``ecological crisis'' stage. We will provide a detailed explanation of how EI affects the Seaweed Swarm System and the Fungi System.

\renewcommand{\thefigure}{2}
\begin{figure}[H]
\includegraphics[width=3.31in]{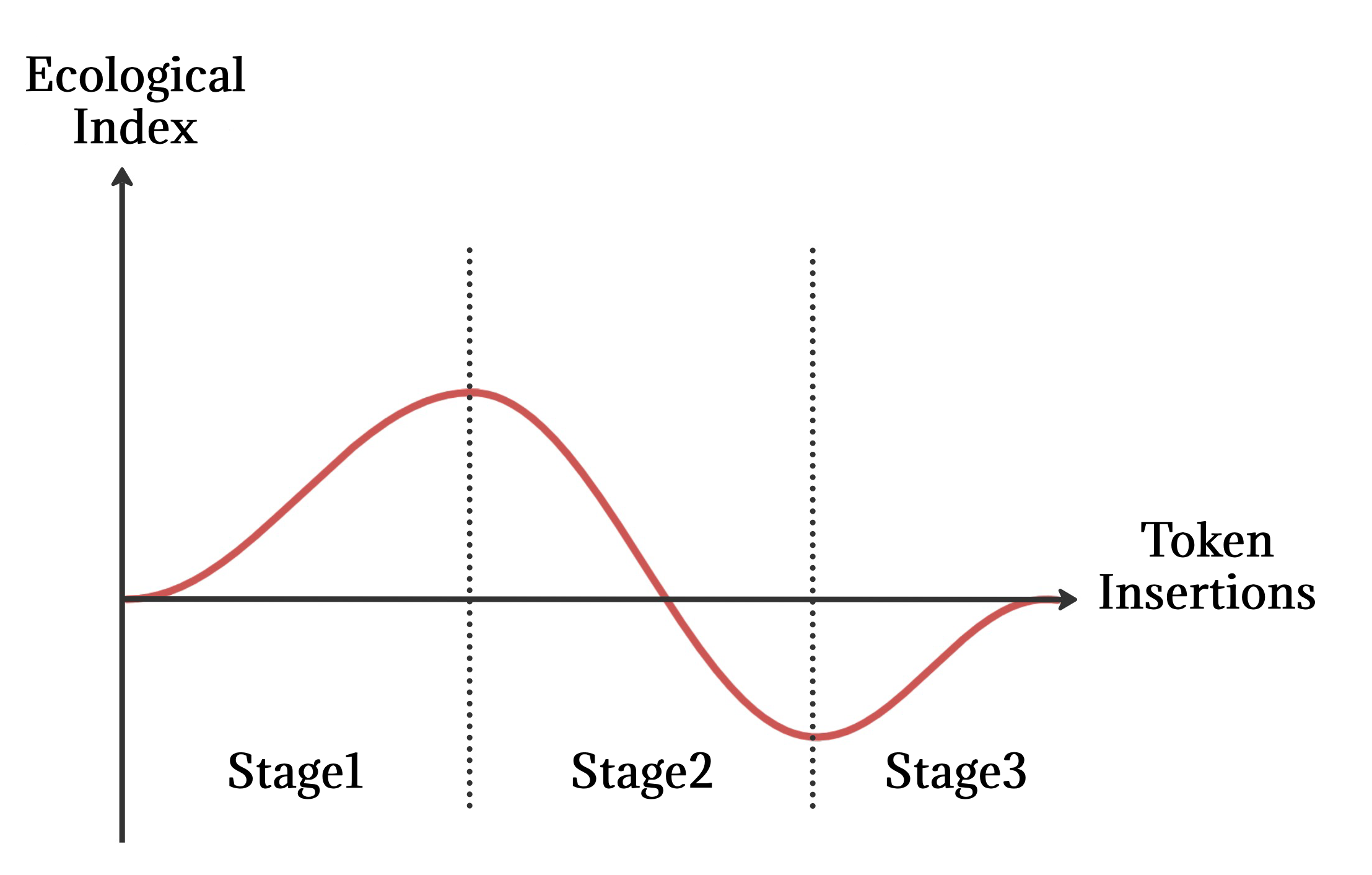}
\caption{Complete cycle of the relationship between token insertions and ecological index. \copyright Danlu Fei and Ziwei Wu.}
\label{fig:ei}
\end{figure}

\begin{itemize}
  \item \textbf{Seaweed Swarm System and EI:}

EI is directly proportional to the growth rate of seaweed in the seaweed swarm. In the ``prosperity of growth'' stage, human interactions positively influence the ecology and promote seaweed growth. Seaweed grows faster as tokens are inserted. In the ``decline of growth'' stage, human interactions disrupt the ecology, resulting in a slowdown of seaweed growth until it stops completely. When EI reaches the ``ecological crisis'' stage, seaweed growth completely stops, but token insertions continue to harvest seaweed. If the remaining number of seaweed is sufficient, continued token insertions can gradually restore EI to the starting point of the next cycle, representing ecological restoration. Otherwise, continued human interactions may lead to the complete extinction of the entire seaweed swarm, a severe consequence of excessive harvesting. In this situation, the audience still needs to keep inserting tokens to restore EI to the starting point of the next cycle. However, since all seaweed has gone extinct, these token insertions will not harvest any seaweed and will not yield any profits. This is designed as ecological repayment.

In addition to affecting seaweed growth rate, EI also influences five natural factors~(also defined as global variables) in the ecosystem: water temperature, salinity, flow velocity, irradiation, and nutrient concentrations. These factors determine the shape of newly grown seaweed in real-time~(Fig.~\ref{fig:ml}), which further decides the selling price of the seaweed. Details on this aspect will be discussed in the Technical Implementation section when introducing Seaweed Generation.
  
  \item \textbf{Fungi System and EI:}

Although in our design, audience token insertions for the Fungi System do not affect EI, EI can influence the Fungi System. As aforementioned, each time the audience inserts a coin for the Fungi System, a fungus can be cultivated to treat diseases caused by oomycetes, which increases the selling price of seaweed. The curing difficulty is inversely proportional to the EI. The better the ecological condition with a larger EI, the fewer fungi needed to kill the oomycete, and vice versa.

Meanwhile, the killed oomycete periodically respawns, re-infecting the seaweed swarm. The oomycete's respawn rate is inversely proportional to EI too. A smaller EI indicates worse ecology, leading to faster oomycete respawning and a quick re-infection of the previously healthy seaweed swarm.
\end{itemize}

\section{Technical Implementation}
\subsection{Seaweed Generation}
As aforementioned, in our Benefit Game, the shape of each seaweed directly determines its selling price. Therefore, instead of using random functions or self-defined rule-based methods to determine the shape of seaweeds, we believe that it is more appropriate to train a machine learning model using real-world seaweed aquaculture data and PCGML to determine the shape and generate seaweed. Our training set includes data collected from research on seaweed aquaculture in the real world~\cite{Corey2014,Tresnati2021,Kregting2016,Xiao_2019}, which demonstrate the effects of five natural factors~(water temperature, water salinity, water flow velocity, irradiation, and nutrient concentrations) on seaweed yield. Notably, our dataset only includes the relationship between natural factors and seaweed yield, and not seaweed shape. After training the model, we map the output~(seaweed yield) to the seaweed shape to generate various seaweed variants~(Fig.~\ref{fig:variants}).

\renewcommand{\thefigure}{4}
\begin{figure}[t]
\centering
\includegraphics[width=2.8in]{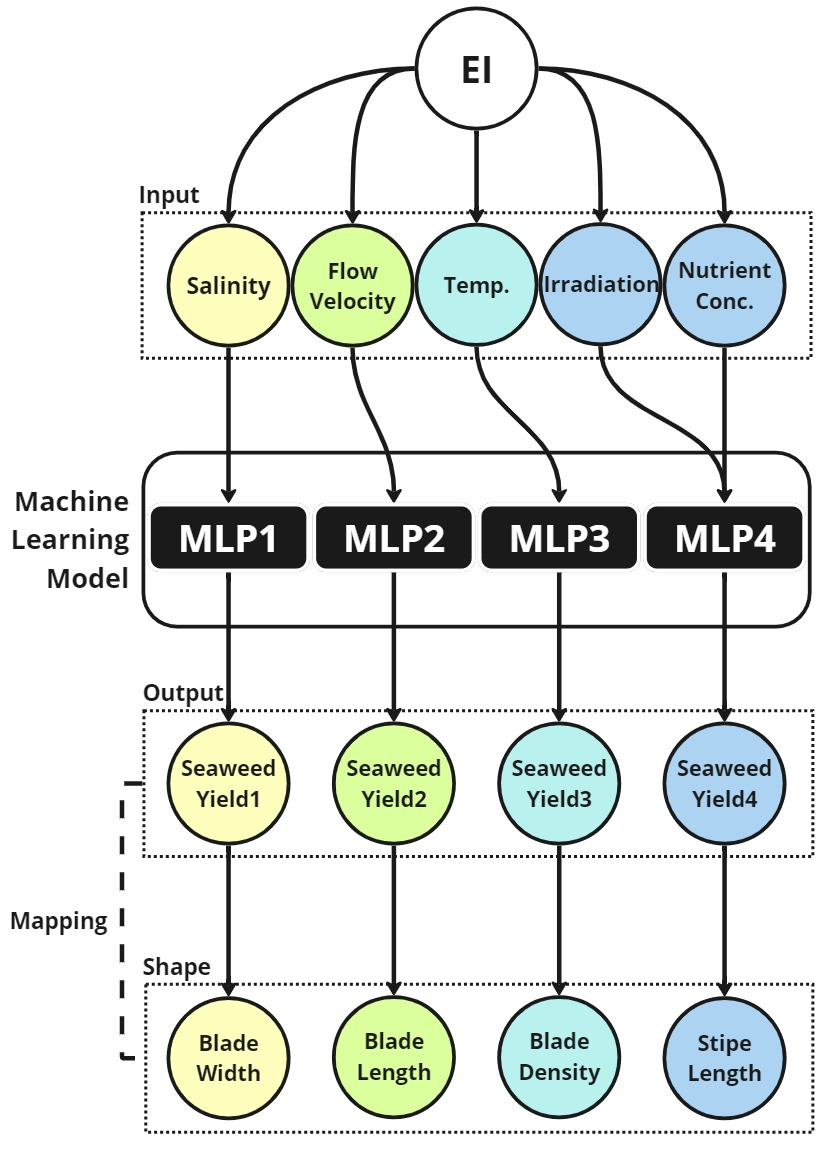}
\caption{Overall structure of machine learning model used for seaweed generation. \copyright Danlu Fei and Ziwei Wu.}
\label{fig:ml}
\end{figure}

\renewcommand{\thefigure}{5}
\begin{figure}[ht]
\includegraphics[width=3.31in]{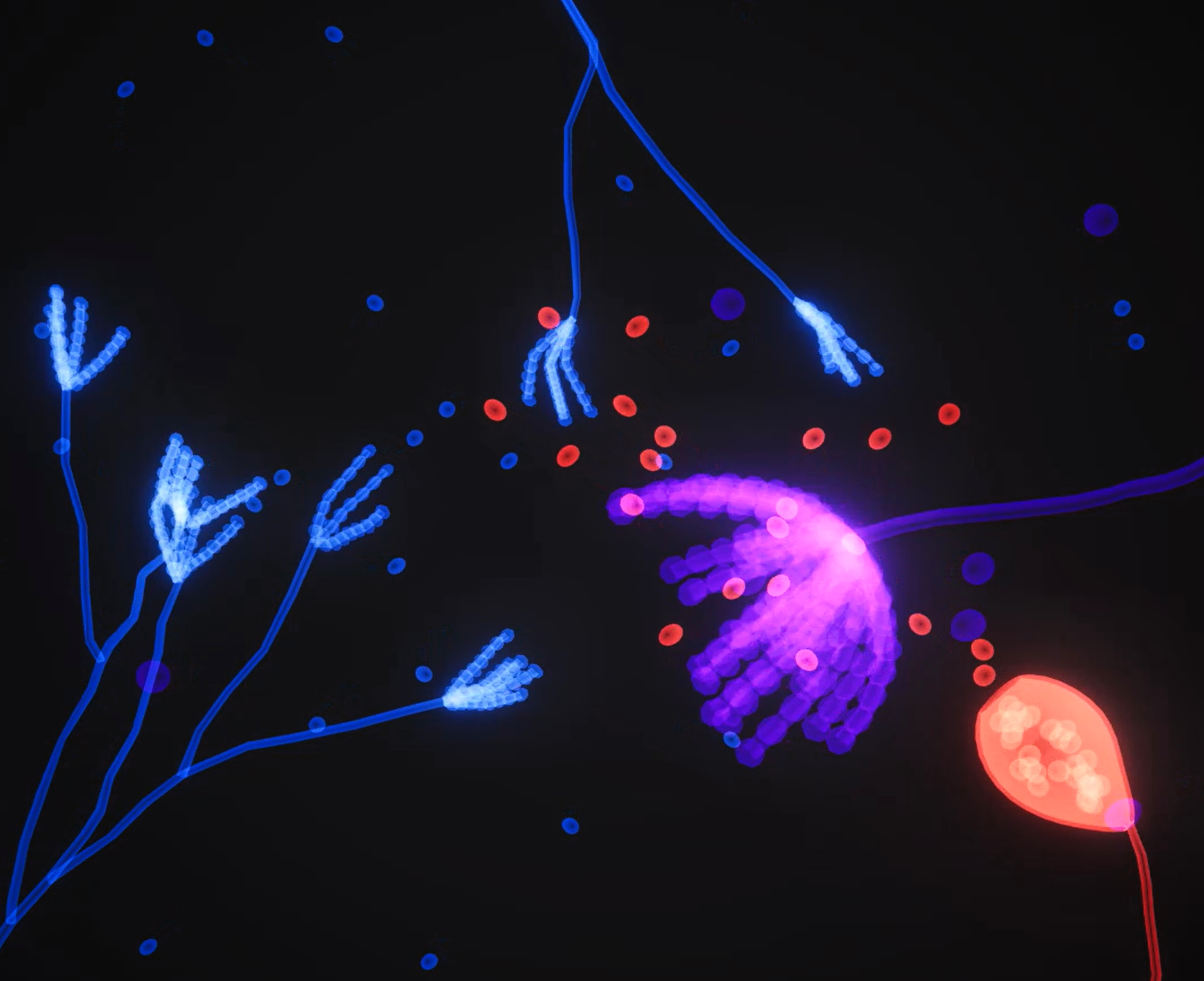}
\caption{Oomycete~(red) and two types of fungi. \copyright Danlu Fei and Ziwei Wu.}
\label{fig:fungi}
\end{figure}

To handle the non-linear relationships in our collected dataset, we choose to use Multilayer Perceptron~(MLP) models. Since our data comes from various sources and is independently associated with the natural factors mentioned earlier, we train separate MLPs for each factor. With the help of the Barracuda package in Unity, we can easily use our trained MLP models within the Unity environment.

After training, we obtain a model with five natural factors as input and four seaweed yields as output~(Fig.~\ref{fig:ml}). Therefore, we set five global variables in the game corresponding to these natural factors as inputs to the model. These variables change with the EI, indicating the impact of EI on the entire seaweed ecology. We map the output of the model to four parameters that determine the seaweed shape: the width of the seaweed blade is determined by salinity, the length of the blade is determined by water flow velocity, the density of the blade is determined by water temperature, and the length of the stipe is determined by irradiation and nutrient concentration. We then transfer the real-time calculated mapped parameters of the seaweed shape to the Houdini plugin in Unity for procedural modeling of seaweed variants with different shapes.

\renewcommand{\thefigure}{6}
\begin{figure*}[t]
\centering
\includegraphics[width=\textwidth]{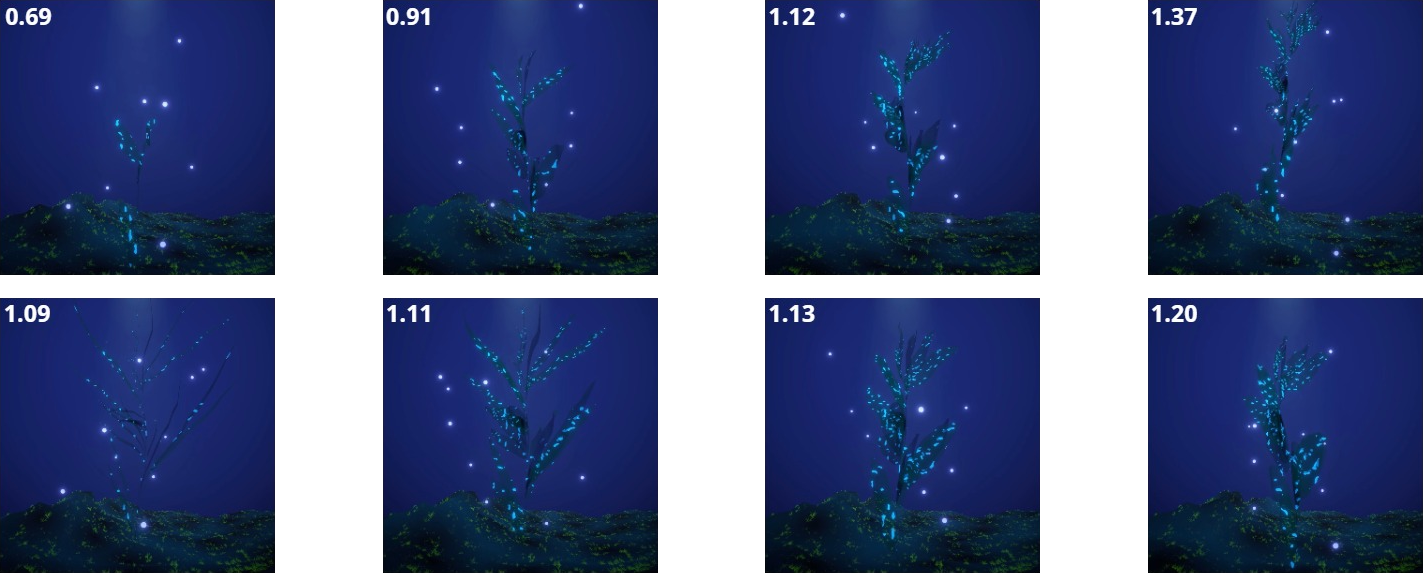}
\caption{Examples of individual seaweed and their corresponding selling prices~(colored in white). In the top row, blade width and length are constants, while stipe length and blade density both increase, from left to right. In the bottom row, stipe length and blade density are constants, while blade width increases and blade length decreases, from left to right. \copyright Danlu Fei and Ziwei Wu.}
\label{fig:variants}
\end{figure*}

\subsection{Fungi Generation}
We procedurally generate two types of fungus inspired by Penicillium~(blue ones in Fig.~\ref{fig:fungi}) and Aspergillus~(purple one in Fig.~\ref{fig:fungi}), which are symbiosis with seaweeds and can treat seaweed diseases caused by oomycetes in the real world~\cite{Vallet2018,Tourneroche_2020}. The cylindrical structures of the stipes, metulae, and phialides of the fungi are initially generated as lines and subsequently assigned thickness and width. The conidia of the fungi are represented directly using spheres. As both fungi resemble tree structures and form branches at different levels, we employ a parent-children structure during generation. The stipe of each fungi is the parent, with metula branches as the children of the stipe, and phialide branches as the children of the metula branches. Each phialide branch also has several spherical conidia as its children. Rules are established for each level of the parent-children structure based on range intervals instead of fixed numbers. For example, the number of child branches can be randomly determined from a range of 2 to 5. As shown in Fig.~\ref{fig:fungi}, one of the blue fungi has only two branches, while the other has five branches. In addition, the length, thickness, angle of branch distribution can also be randomized within range intervals, resulting in the generation of numerous fungi that follow the same rules and have similar structures but unique shapes.


\subsection{Disease Effect}
To simulate rot diseases caused by parasitic oomycetes on the blades of Alien Seaweed, we add irregular glowing patches to each individual seaweed blade. We achieve this effect by writing a fragment shader within Unity's node-based shader editor. We use gradient noise to generate floating-point values ranging from 0 to 1. Subsequently, a step function is employed to map these floating-point numbers to either 0 or 1. Areas that output a value of 0 form the base color of the seaweed, while areas that output a value of 1 form the light patches. This step function has an adjustable parameter called ``edge'', where any floating-point number greater than ``edge'' is mapped to 1, and vice versa for 0. By adjusting ``edge'', we can control the ratio of the base color area to the light patch area, thus simulating seaweed in different health conditions. The healthier the seaweed is, the larger the ``edge'' will be, causing fewer light patch areas~(Fig.~\ref{fig:compare}). Furthermore, the scale of the gradient noise can also be adjusted to control the size of the glowing patches corresponding to the health condition of the seaweed swarm.

\renewcommand{\thefigure}{7}
\begin{figure}[ht]
\includegraphics[width=3.31in]{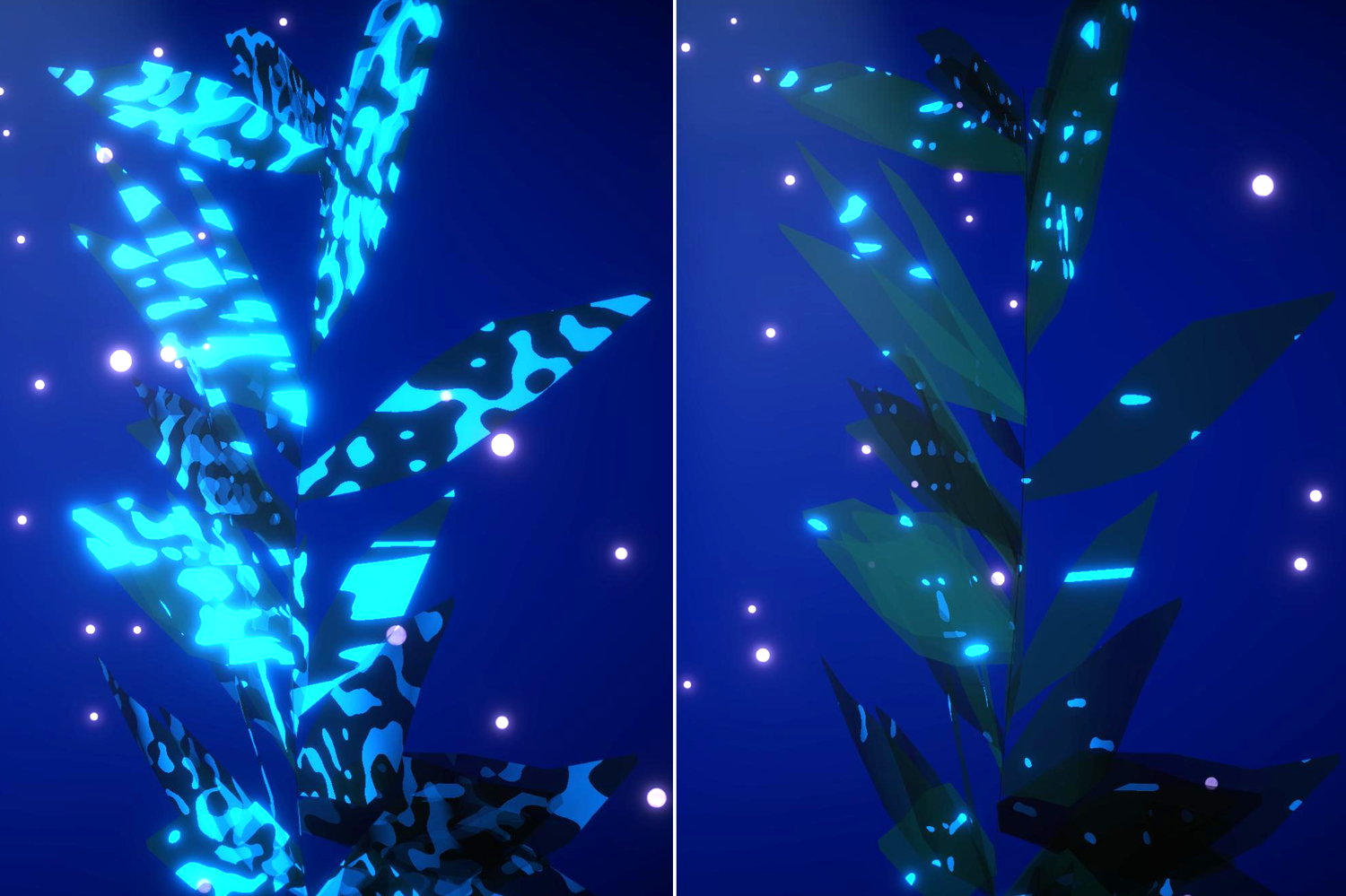}
\caption{Diseased seaweed with large and dense glowing patches~(left), and healthy seaweed with small and loose glowing patches~(right). \copyright Danlu Fei and Ziwei Wu.}
\label{fig:compare}
\end{figure}

\renewcommand{\thefigure}{8}
\begin{figure}[ht]
\centering
\includegraphics[width=3.31in]{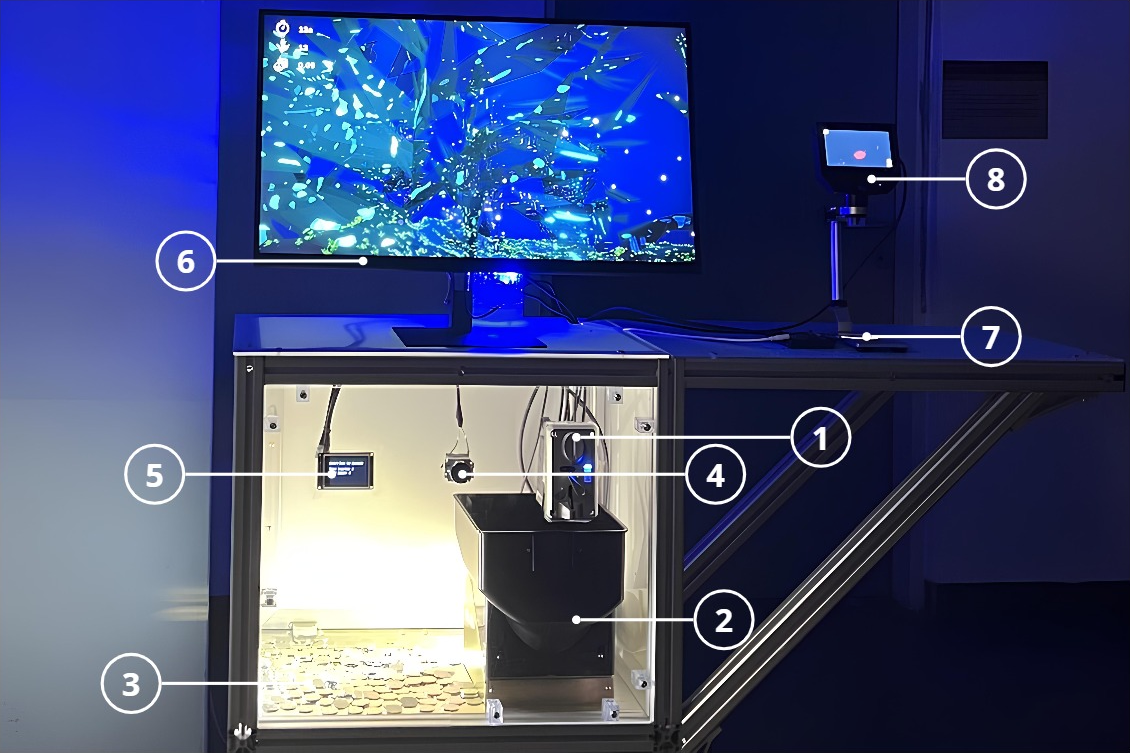}
\caption{Overview of the ``Benefit Game: Alien Seaweed Swarms'' installation.~\copyright Photograph by Bixiang Lu.}
\label{fig:installation}
\end{figure}


\subsection{Installation}
Our game runs on a computer host and all electronic components are controlled by an Arduino board. The frame of the installation is made of aluminum profiles with acrylic sheets as surfaces. It consists of a two-tiered cabinet where the bottom houses the computer host, Arduino board, and power supplies, while the upper serves as a space for audience interaction and game token storage. We add a plane supported by an angled structure at the top of the cabinet to place display devices for the Seaweed Swarm System and the Fungi System separately.
\subsubsection{Token Acceptance}
On the right side of the upper cabinet’s front surface, we install a token acceptor~(No.1 in Fig.~\ref{fig:installation}) to sense valid game token insertion and return non-game tokens to the audience. It will transmit a signal to the Arduino if a valid token is sensed.
\subsubsection{Token Return}
We place a token returner~(No.2 in Fig.~\ref{fig:installation}) below the token acceptor with an open top to catch the dropped tokens. The returner consists of an IR sensor for counting tokens and an electrical motor. When the power is on, the motor rotates continuously and dispenses coins from the side of the returner while the IR sensor counts the number of coins dispensed. We use a relay module to connect the returner's power supply so that when the Arduino emits a high voltage, the returner starts dispensing coins. When the number of coins dispensed recorded by the IR sensor reaches the desired amount, the Arduino emits a low voltage to stop the coin dispensing. The tokens dispensed by the returner are stored in the upper cabinet, and we open a slot~(No.3 in Fig.~\ref{fig:installation}) on the left bottom of the upper cabinet's front panel for the audience to pick up the tokens and continue playing the game.
\subsubsection{Target Switch}
We install a button~(No.4 in Fig.~\ref{fig:installation}) in the middle of the upper cabinet's front panel, enabling the audience to choose whether to insert tokens for the Seaweed Swarm System or the Fungi System by pressing it.
\subsubsection{Arduino and Host Communication}
We utilize serial communication between the Arduino and the computer host. When the Arduino receives a valid coin insertion signal from the token acceptor, it sends a message to the computer host via the serial port, which affects the EI and triggers the harvesting of seaweed or the cultivating of fungi in the game running on the host. When the game requires a profit settlement for the harvested seaweed every 20 seconds, the computer host sends a message to the Arduino to activate the relay module with a high voltage, thus running the returner.
\subsubsection{Game Display}
The installation includes three display devices. On the left side of the upper cabinet's front panel, we have a 2.4-inch TFT screen~(No.5 in Fig.~\ref{fig:installation}) that displays the current coin insertion target, as well as the total number of coins inserted and dispensed. The Seaweed Swarm System uses a conventional monitor~(No.6 in Fig.~\ref{fig:installation}) located at the top of the cabinet. The monitor features a user interface comprised of icons that provide information such as the countdown for token settlement, the current remaining quantity of seaweed, and the unsettled profits of the harvested seaweed. To simulate the effect of observing fungi under a microscope, we use a circular screen~(No.7 in Fig.~\ref{fig:installation}) the size of a petri dish for the display of the Fungi System, and an electronic magnifying glass~(No.8 in Fig.~\ref{fig:installation}) to enlarge the display content. The display device for the Fungi System is located on the extended plane at the top of the cabinet.


\section{Audience Reaction}
``Benefit Game: Alien Seaweed Swarms'' was exhibited at a major exhibition  in Linz in 2023. This exhibition provided an opportunity to test our installation, observe the audience's reaction, and collect feedback on their interactions~(Fig.~\ref{fig:audience}). We received many positive reviews during this exhibition, and accordingly, we also identified issues that need to be addressed.


For example, children were likely keen on harvesting seaweed by continuously inserting tokens into the installation. The activity resulted in a quick profit, but also caused immediate damage to the ecosystem, which will require time for the environment to recover. During this waiting period, the children quickly became bored and wanted the seaweed to grow faster. The vast majority of the participants enjoyed the feedback of earning money, and many would stay in front of the installation for a long time. Additionally, it is interesting to note that some participants were obsessed with inserting tokens for the Fungi System, hoping to kill the oomycete. They were happy when enough fungi were cultivated to kill the oomycete but soon became annoyed once a new oomycete grew. Many participants were interested in the story behind the seaweed and tried to connect it to human activities throughout history, such as the development of seaweed in skincare products and food.

After communicating with the audience, we found that most of them believed that our game increased their awareness of the importance of ecological sustainability and encouraged them to find a balance in their interactions. Some witnessed the ecosystem crisis caused by over-harvesting, leading to the extinction of the seaweed ecologies, and felt a sense of loss and a desire to restore the ecosystem to a healthy state as soon as possible. Some even spontaneously advised others who blindly harvested seaweed for profits to consider the impact of their behavior on the current seaweed growth rate, remaining seaweed quantity, and ecological conditions.

We received feedback from the audience who found it not intuitive to switch interaction targets and use the same token acceptor for two systems. Although we display the current interaction target on the TFT screen, the audience often overlooks it and relies on observing which system responds after token insertion. Also, some participants wished us to present more information about the current ecological state, such as a real-time display of the seaweed growth rate curve, to help them interact sustainably with the seaweed ecosystem.

Through gameplay, we aim to foster a deeper understanding of the Capitalocene and Anthropocene among the audience from a first-person perspective, prompting players to question and reflect on the impact of their behavior on economic growth and environmental stewardship.

\renewcommand{\thefigure}{9}
\begin{figure}[H]
\includegraphics[width=3.31in]{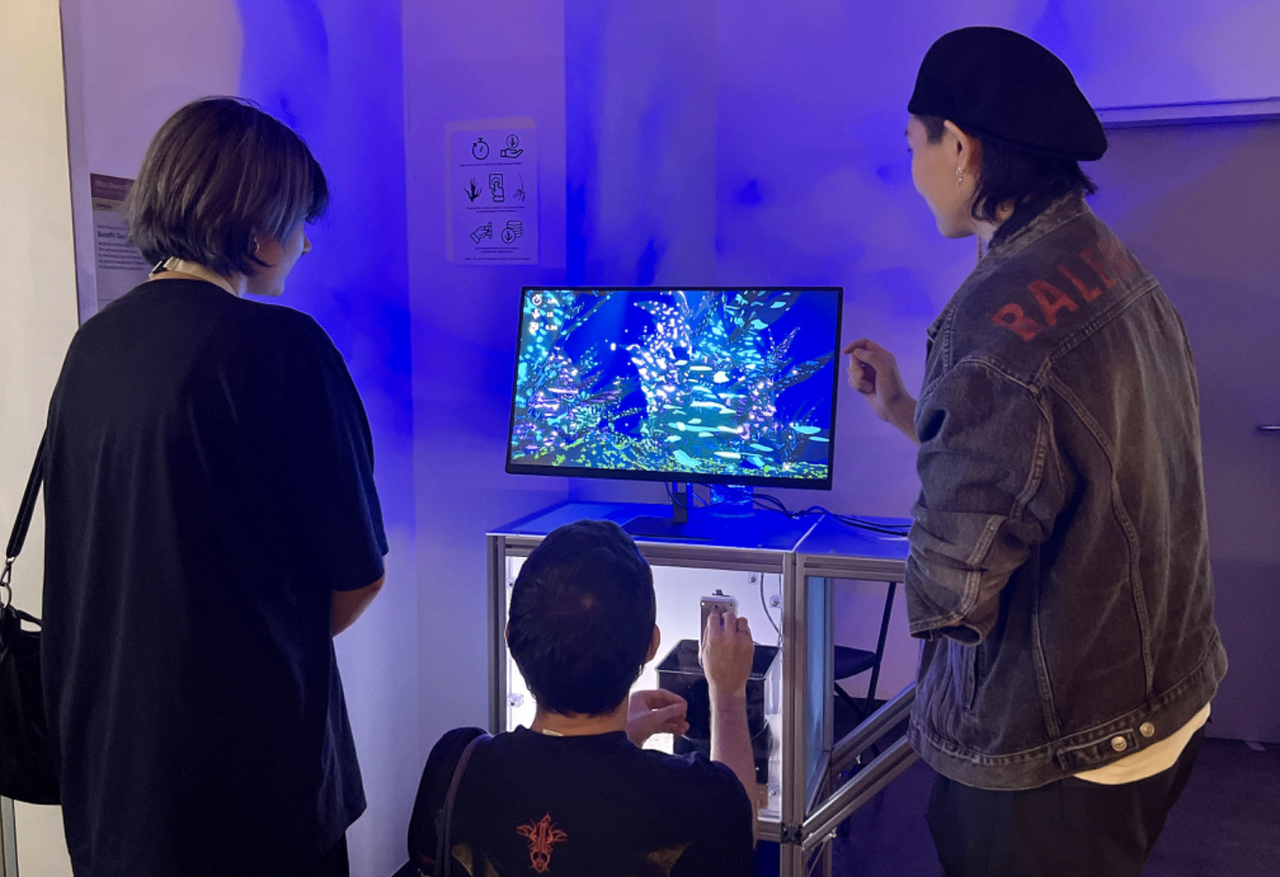}
\caption{Audience interacting with the installation at the exhibition. \copyright Photograph by Bixiang Lu.}
\label{fig:audience}
\end{figure}

\section{Conclusion and Future Work}
We have developed an ecological game that creates a virtual environment of the seaweed ecosystem. Our work utilizes PCGML to generate seaweed, which is based on data related to natural factors and seaweed yield in the real world, and mapped to the shape of seaweed. The generation of fungi also employs rule-based PCG technology. Our aim is to raise ecological consciousness among players. Players can interact with the entire ecosystem using game tokens. Inserting tokens for the Seaweed Swarm System allows players to harvest seaweed and earn game tokens back as profits. On the other hand, inserting tokens for the Fungi System allows players to cultivate the symbiosis fungi that can treat seaweed diseases caused by the oomycete, increasing the seaweed selling price. Players need to consider how to interact with the ecosystem sustainably, as over-harvesting can have a negative impact on the ecosystem and even lead to ecological crises that require significant effort to recover. Meanwhile, ignoring seaweed diseases in pursuit of profits would also lower the seaweed selling price, resulting in a counterproductive outcome. Therefore, we hope that players will explore and find a balance point that enables sustainable development of the ecosystem.

In the future, we plan to present the real-time status of the ecosystem, as well as the input and output of tokens, and present them on a separate screen for the audience to track the changes in the ecosystem and the corresponding previous interactions. This will enable the audience to make better decisions on how to interact with the ecosystem for sustainability. Additionally, we plan to update our exhibition methods. On the hardware side, we plan to increase the current token acceptor from one to two, which will control the token input for the Seaweed Swarm and Fungi System respectively, making the interaction target more intuitive for the audience. We also plan to use projection instead of our current monitor to provide the audience with a more vivid sense of the size of the seaweed swarm and enhance the visual effect of the exhibition.



\bibliographystyle{unsrt}
\bibliography{isea}


\end{document}